\documentclass[useAMS,usenatbib,usegraphicx]{mn2e}

\usepackage{psfig}
\usepackage{times}
\usepackage{subfigure}

\newcommand{\kms}{\hbox{km s$^{-1}$}}
\newcommand{\ms}{\hbox{m s$^{-1}$}}

\newcommand{\vsini}{\hbox{$v$\,sin\,$i$}}

\newcommand{\degs}{$\degr$}
\newcommand{\chisq}{$\chi^{2}$}

\newcommand{\dom}{\hbox{$\Delta\Omega$}}

\begin{document}

\title[Differential Rotation]{The Dependence of Differential Rotation on Temperature and Rotation}

\makeatletter
 
\def\newauthor{%
  \end{author@tabular}\par
  \begin{author@tabular}[t]{@{}l@{}}}
\makeatother

\author[J.R.~Barnes, A.~Collier~Cameron, J.-F.~Donati, D.J.~James, S.C.~Marsden, P.~Petit]
{J.R.~Barnes$^1$\footnote{E-mail: jrb3@st-andrews.ac.uk} A. Collier Cameron$^1$,  J.-F.~Donati$^2$, D.J.~James$^3$, S.C.~Marsden$^4$, P.~Petit$^5$ \\
$^1$ School of Physics and Astronomy, University of St Andrews, Fife KY16 9SS. UK. \\ 
$^2$ Laboratoire d'Astrophysique, Observatoire Midi-Pyrénées, F-31400 Toulouse, France \\
$^3$ Department of Physics and Astronomy, VanderBilt University, Nashville TN 37235, USA \\
$^4$ Institute of Astronomy, ETH Zentrum,  CH-8092, Zurich, Switzerland \\
$^5$ Max-Planck Institut für Aerononomie, Max-Planck-Str. 2, 37191 Katlenburg-Lindau, Germany \\
}

\date{2004, 2004}

\maketitle

\begin{abstract}

{ We use Doppler imaging techniques to determine the dependence of starspot rotation rates on latitude in an homogeneous sample of young, rapidly-rotating solar analogues. A solar-like differential rotation law is used, where the rotation depends on sin$^2$($\theta$), where $\theta$ is the stellar latitude. By including this term in the image reconstruction process, using starspots as tracers, we are able to determine the magnitude of the shear over more than one rotation cycle. We also consider results from matched filter starspot tracking techniques, where individual starspot rotation rates are determined. In addition we have re-analysed published results and present a new measurement for the K3 dwarf, Speedy Mic. 

A total of 10 stars of spectral type G2 - M2 are considered. We find a trend towards decreasing surface differential rotation with decreasing effective temperature. The implied approach to solid body rotation with increasing relative convection zone depth implies that the dynamo mechanism operating in low-mass stars may be substantially different from that in the Sun.}

\end{abstract}


\section{INTRODUCTION}

\begin{figure}
\begin{center}
\includegraphics[angle=0,width=86mm]{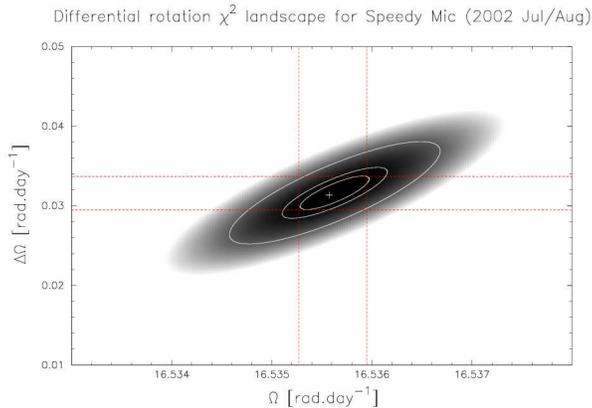} \\
\end{center}
\caption[]{The figure here shows a \chisq~landscape for the 2002 July 18, 19 \& 23 Speedy Mic (HD 197890) timeseries, revealing that rotation rate and differential rotation are correlated. The minimum point (marked by a cross) gives the best fitting differential rotation and related errors through the 1-parameter 1-$\sigma$ confidence interval (inner contour). Also shown are the 2-parameter 1-$\sigma$ and 2.6-$\sigma$ (99\%) confidence intervals.}
\protect\label{fig:speedy}
\end{figure}

{ \citet{carrington1860} was the first to attribute the scatter in rotation rates of sunspots to a differential rotation of the solar surface. It would however be a further 50 years before \citet{hale1908} would discover the existence of magnetic fields in Sunspots.} Helioseismological studies of acoustic-wave propagation through the solar interior confirm that the equatorial regions rotate more rapidly than the polar regions throughout the outer convective zone (e.g. \citealt{schou98helio}). Radial shear is most pronounced at the tachocline which is located at the interface between the convection zone and radiative solar interior. 

The tachocline provides a relatively stable environment for the amplification of magnetic field, being located within the region where convective cells can sink and overshoot into the solar core. { In this region of strong radial shear, the $\Omega$-effect stretches poloidal (North-South field) field into toroidal (East-West) field.} Toroidal field components entrained in convecting fluid elements rotate under the influence of Coriolis forces as they rise and fall, regenerating the poloidal field via the $\alpha$-effect { \citep{parker55,krause80}}. { One} goal of solar dynamo theory is to combine these effects into a self-consistent physical model that can reproduce the 11-year solar activity cycle in detail.

Magnetic activity in stellar counterparts of the Sun gives rise to periodic optical variability as starspots rotate into and out of view. { This variability is also seen in optical and UV emission lines from the chromosphere, and in the coronal X-ray luminosity.} All these indicators are correlated with photospheric magnetic field strength on the Sun, and increase with rotation rate in stars of a given mass. 

Long term photometric monitoring of starspot modulation in magnetically-active binary stars by \citet{hall91dynamo} and \citet{henry95diffrot} (H95) has revealed secular changes in rotation period which are strongly indicative of rotational shearing. As a stellar activity cycle progresses, starspots should emerge at different latitudes. If the star rotates differentially, these changes in latitude will produce long-term scatter in the photometric rotation periods { measured at single epochs}. The correlation between this scatter and the mean rotation period indicates that $\Delta\Omega \propto {  \Omega^{0.24 \pm 0.06}}$  (where $\Omega$ is angular velocity and \dom~the corresponding scatter in radians day$^{-1}$), { and thus} that \dom~is almost independent of rotation {(H95)}.

The Mount Wilson Survey of chromospheric \hbox{Ca {\sc ii} H \& K} emission-line fluxes has shown that many solar-like stars exhibit multi-periodic chromospheric emission flux variations \citep{wilson78,baliunas95}. The short-period modulation (of order days to weeks) gives the rotation period of the dominant active latitude at any epoch. The longer-period modulations correspond to the solar activity cycle. Over timescales of many years, the rotation period is itself found to show a characteristic sinusoidal variation in many stars of spectral type F to K. Cyclic changes in the period are attributed to a stellar dynamo driven by differential rotation. { \citet{donahue96} (DBS96) have shown that $\Delta\Omega \propto \Omega^{0.7 \pm 0.1}$, a result which is not consistent with the findings of H95, even within the quoted uncertainties. The sample of H95 comprises mostly binary systems, although a correction term for filling factor is included in their analysis.  It should be emphasised here that no latitudinal information about the starspot groups or active regions is assumed in the $\Delta\Omega/\Omega$ proportionality. A more significant explanation for the discrepancy of the results presented by H95 and DBS96 is related to this issue and will be discussed in \S \ref{section:discussion}.}

\section{Method \& Results}
\protect\label{section:results}

\begin{table}
\caption{{ Colours, temperatures and differential rotation measurements.} References in column 6 are: 1) \citet{marsden04r58}, 2) \citet{marsden04gdwarfs}, 3) \citet{donati00rxj1508} 4) \citet{barnes00pztel}, 5) \citet{cameron02twist}, 6) \citet{donati03temporal}, 7) This paper, 8) \citet{barnes04lopeg}, 9) \citet{barnes04hkaqr}. References marked * denote re-calculated \dom~using the sheared image method (see text for details).}
\protect\label{tab:table}
\vspace{5mm}
\begin{center}
\begin{tabular}{lccccc}
\hline
Object			& B-V	& T			& \dom				& $\Delta\Omega_{err}$	&	Ref. \\
				&		&	[K]		& [rad.day$^{-1}$]		&					&		 	  \\
\hline

R58			 &	0.65 	&	5859	&		 0.02500	&	  0.01500	&	   1 \\
R58 			&	0.65 	&	5859 	&		 0.14000	&	  0.01000	&	   2 \\
(HD307938)	&			&			&				&				&		\\
LQ Lup 		&	0.69		&	5729 	&		 0.13000	&	  0.02000	&	   3 \\
(RX J1508.6-4423) &		& 			&				&				&		\\
PZTel 		&	0.78 	&	5448 	&		 0.10134	&	  0.00654	&	   4* \\
VXR54A		&	0.81 	&	5170 	&		 0.07000	&	  0.02000	&	   2 \\
ABDor1 		&	0.82 	&	5386 	&		 0.04620	&	  0.00578		&	   5 \\
ABDor2 		&	0.82 	&	5386 	&		 0.09106	&	  0.01188	&	   5 \\
ABDor3 		&	0.82 	&	5386 	&		 0.08850	&	  0.00748		&	   5 \\
ABDor4 		&	0.82 	&	5386 	&		 0.06684	&	  0.02027		&	   5 \\
ABDor5 		&	0.82 	&	5386 	&		 0.07140	&	  0.00568	&	   5 \\
ABDor6 		&	0.82 	&	5386 	&		 0.05764	&	  0.00476	&	   5 \\
ABDor7 		&	0.82 	&	5386 	&		 0.05340	&	  0.00250	&	   6 \\
ABDor8 		&	0.82 	&	5386 	&		 0.04710	&	  0.00250	&	   6 \\
ABDor9 		&	0.82 	&	5386 	&		 0.05840	&	  0.00150	&	   6 \\
ABDor10 	&	0.82 	&	5386 	&		 0.04610	&	  0.00280	&	   6 \\
ABDor11 	&	0.82 	&	5386 	&		 0.05400	&	  0.00130	&	   6 \\
Speedy Mic	&	0.93 	&	4989 	&		 0.03157	&	  0.00206	&	   7  \\
(HD 197890)	&			&			&				&				&		\\
LQHya1 		&	0.92 	&	5019 	&		 0.19420	&	  0.02160	&	   6 \\
LQHya2 		&	0.92 	&	5019 	&		 0.01440	&	  0.00290	&	   6 \\
LOPeg 		&	1.08 	&	4577 	&		 0.03550	&	  0.00682	&	   8 \\
HKAqr 		&	1.42 	&	3697 	&		 0.00496	&	  0.00917	&	   9 \\
EYDra 		&	1.45 	&	3489 	&		 0.00030	&	  0.00333		&	   7* \\

\hline

\end{tabular}
 \end{center}
\end{table}

The work reported here employs the Doppler imaging technique, which offers a more direct approach { than the above methods} to measuring stellar surface rotation rate as a function of latitude. This sheared image method allows us to make an indirect image of the stellar surface using a model which includes the effects of differential rotation and has been investigated by \citet{petit02}. The image-data transformation incorporates a simplified solar-like differential rotation law, where rotation is proportional to the square of the sine of the stellar latitude, \mbox{$\theta$, i.e. $\Omega(\theta) = \Omega_{eq} - \dom.$sin$^{2}(\theta)$}. We determine the value \dom~of the stellar surface shear that gives the best fit to the data secured over several stellar rotations. Since spots are present at a variety of latitudes, their individual rotation rates trace the differential rotation pattern. { Fig. \ref{fig:speedy} demonstrates how the goodness of the final fit to the data is used to determine the equatorial rotation rate. Here we present the results from data taken with the Anglo Australian Telescope and the University College London \'{E}chelle Spectrograph. Observations of the 0.38 d period K3 dwarf Speedy Mic (HD 197890) were secured on the nights of 2002 July 18, 19 \& 23 and will be presented in detail in a separate publication.}

Additionally, \citet{cameron02twist} applied a matched filter analysis technique, allowing the rotation rates of individual starspots to be determined { directly from the timeseries spectra}. We include those measurements for AB Dor in addition to results from the sheared image method \citep{donati03temporal}.  
\begin{figure*}
\begin{center}
\includegraphics[angle=270,width=14cm]{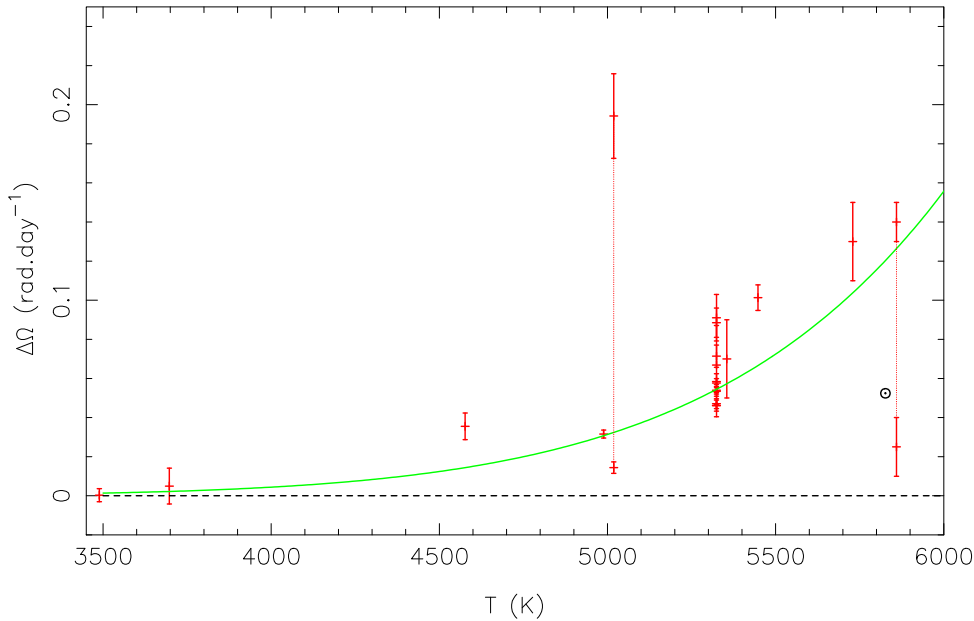} \\
\end{center}
\caption[]{Differential rotation (\dom) as a function of stellar temperature. The circled dot represents the Sun. Independent measurements, made in different years, are shown for three of the stars in our sample: AB Dor, LQ Hya and R58.  A dashed line connects pairs of points for LQ Hya and R58.}
\protect\label{fig:domega_temp}
\end{figure*}

This technique has been applied to an homogeneous sample of ten young, rapidly rotating stars with spectral types ranging from G2V to M2V. The measured shear, B-V colours and references are listed in Table \ref{tab:table}. The model-atmosphere relation given by \citet{bessell98colours} was used to obtain the effective surface temperature from the observed  B-V colours. In two cases, we have re-determined \dom~using the sheared image method. For PZ Tel \citep{barnes00pztel} we initially used image cross-correlation which yielded a slightly higher value for the latitude dependent shear. We have also re-analysed the results from \citet{barnes01mdwarfs} for the M1-2V EY Dra (RE 1816+541). In this latter instance, we initially used splines to interpolate the grid search of \dom~vs $\Omega$~which gave a very non-uniform \chisq~landscape with more than one minimum. We now fit a quadratic function in both dimensions eliminating local fluctuations and yielding a single global minimum. The combined results from all studies are plotted in Fig. \ref{fig:domega_temp}, and reveal a marked trend towards lower differential rotation with decreasing stellar effective temperature. We find a power-law dependence of the surface shear on stellar surface temperature: 

\begin{equation}
\Delta\Omega \propto T^{8.92 \pm 0.31}
\end{equation}

{ Those stars with more than one measurement of the surface shear show} that the magnitude of the scatter is greater than that of the uncertainty on a single measurement, indicating secular variability \citep{cameron02twist, donati03temporal} in either the differential pattern itself or in the depths at which the fields giving rise to the tracers are anchored \citep{donati03temporal}. Nevertheless LQ Hya poses the most problematic point in the measured relation. Careful analysis suggests that the measured shear values are reliable. We note that LQ Hya is the slowest rotator in the sample, yielding relatively poor spatial resolution ($\sim$16\degs) at the stellar equator. Short-term evolution of unresolved spot groups may lead to a systematic mis-determination of the differential rotation. We discuss this issue and secular variations of \dom~further in \S \ref{section:discussion}.

\begin{figure*}
\begin{center}
\includegraphics[angle=270,width=14cm]{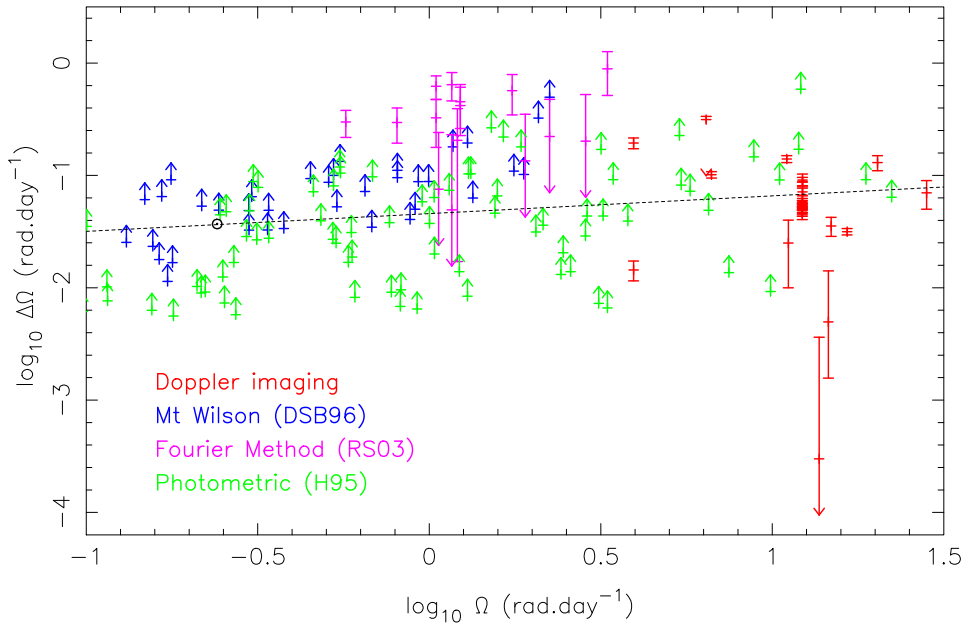} \\
\end{center}
\caption[]{Differential rotation (log10(\dom)) as a function of  log10($\Omega$). Results from various studies: Doppler imaging (see Table 1), DSB96, H95 and RS03. The Sun is represented by the circled dot. The dashed line is an unweighted fit to all points with slope \mbox{$0.15 \pm 0.10$}.}
\protect\label{fig:domega_omega}
\end{figure*}

Fig. \ref{fig:domega_omega} shows the dependence of differential rotation, \dom, on equatorial rotation rate, $\Omega$, for the stars in our sample, along with available data from other studies. { We discuss the results of \citet{reiners03diffrot} (RS03) in \S \ref{section:discussion}. The unweighted least squares fit (i.e. neglecting errors or limits) yields a slope indicating that \dom\ $\propto \Omega^{0.15 \pm 0.10}$. The values for \dom~derived from Doppler imaging studies and the solar value (circled dot) represent the full difference in rotational frequency between the pole and the equator, while the values for \dom~derived from DSB96 and the H95 are lower limits. In other words, they represent the spread in rotation rates only over the range of latitudes where active regions erupt during the stellar cycle. The scatter in \dom~at any given rotation rate is significantly greater than the trend in mean \dom~over the range of $\Omega$ plotted.  

Despite the limitations of an unweighted fit, the power law relationship between \dom\ and $\Omega$ is consistent with the findings of H95. Here we can see that the slope in the DSB96 data alone is greater. A possible explanation is that there is a spectral type dependency on the range of latitudes at which magnetic flux errupts. The faster rotators in the DSB96 sample tend to be later spectral type, while the slower rotators are of earlier spectral type. The latter results are in agreement with RS03 who obtained significant measures of differential rotation for early G dwarfs and F dwarfs only. Despite the high degree of scatter, there is some evidence for spectral type dependence in Fig. \ref{fig:domega_omega} when we compare the results of RS03 (high \dom) with those presented for the M dwarf stars using Doppler imaging studies (low \dom). It is this dependence which we have discerned for the first time in Fig. \ref{fig:domega_temp} for a narrow range of $\Omega$.}

\section{Discussion}
\protect\label{section:discussion}

RS03 found recently that main-sequence stars of spectral types F and G  exhibit  significantly greater differential rotation than we find in the sample of lower-mass stars presented here. The high rotational shear of the F and G stars manifests itself in the Fourier transforms of their rotationally broadened line profiles, allowing precise measurements of shear from line profiles alone for stars with intermediate rotation rates. { We have discussed the enhanced rotational shear (Fig. \ref{fig:domega_omega}) in the thin convective zones of these stars in \S \ref{section:results}.}

\citet{kitchatinov99drot} have modelled internal fluid motions to obtain the surface differential rotation in various stellar types. Their model predicts that the surface differential rotation rate should vary only weakly with axial rotation period and spectral type in the G2 - K5 range. A K5 star is predicted to show approximately 2/3 the surface shear of a G2 star. This is in qualitative if not quantitative agreement with our finding that differential rotation is an order of magnitude lower at K5 than at G2. The stars in our sample rotate more rapidly that the fastest stars in the models of Kitchatinov \& R\"{u}diger, and it is not clear what effects such rapid rotation would impose on model solutions. These authors however found that \dom~$\propto \Omega^{0.15}$ on average for a G2 star and \dom~$\propto \Omega^{-0.05}$ for a K5 star, in reasonable agreement with the unweighted fit (\dom~$\propto \Omega^{0.15 \pm 0.10}$) shown in Fig. \ref{fig:domega_omega}.

That scatter is seen for some stars with more than a single measurement requires further comment. { Being the brightest member of its class,} AB Dor is the most studied single rapid rotator. While there is a slight offset between \dom~as measured using starspot tracking and the sheared image techniques, the overall seasonal variations show that the  equator-pole lap time varies from \hbox{70 d to 140 d}. An analogue of the torsional oscillations seen at the stellar surface \citep{howard80} and within the convection zone \citep{kosovichev97} may be responsible, but the variation in shear is between one and two orders of magnitude greater than the solar variations which amount to 5\ms~deviations from fixed differential rotation. It has been shown \citep{applegate92} that the transport of angular momentum may change in response to changes in the magnetic flux threading the convection zone. This behaviour may be expected to be periodic if magnetic activity cycles are present. \citet{cameron02twist} discuss these possibilities and show that the observations of AB Dor are consistent with the mechanism put forward by Applegate. Alternatively, \citet{donati03temporal} { have} suggested that the starspots which trace differential rotation { may be} anchored at different depths in the convection zone. This implies { that} the magnetic field is generated through the action of a distributed dynamo in rapid rotators.
 
In the case of the more slowly rotating LQ Hya however, we see more than an order of magnitude variation in \dom. We have briefly suggested (\S \ref{section:results}) a possible cause of mis-determination of differential rotation shear in more slowly rotating stars, but note this argument does not apply to the rapidly rotating R58 \hbox{(\vsini = 92 \kms)} which shows considerable variation in shear. Clearly multi-epoch measurements of differential rotation are required for each star if this phenomenon is to be investigated further. Our confidence in the \dom~vs temperature trend is affirmed by the observation that a statistical sample of both M dwarfs and mid-late K dwarfs all exhibit a small \dom~while the findings of \citet{reiners03diffrot} for F dwarfs indicate greater \dom~than for those stars in our sample. 

The decrease in differential rotation that we observe toward later spectral types suggests that the importance of the $\alpha$-effect relative to the $\Omega$-effect  should increase strongly toward lower-mass stars. The $\alpha$-effect should also increase with increasing rotation rate, since it arises from the action of  Coriolis forces on a convecting fluid { \citep{krause80}}. The rotational shear, however, appears {to be weakly dependent on} rotation rate, and indeed decreases strongly towards lower stellar masses. We conclude that the $\alpha$-effect is therefore likely to dominate stellar magnetic-field generation in low-mass stars, and in rapidly rotating young stars with an even wider range of stellar masses { (e.g. \citealt{kuker99alpha})}

%For an homogeneous sample of young rapidly rotating stars, we have shown that the spectral type dictates the variation of \dom~and not $\Omega$ itself. This implies that the importance of the $\alpha$-effect increases towards later spectral types. This may be especially relevant in rapid rotators since the $\alpha$-effect, the effective twisting of emerging field, is driven by Coriolis forces. With no shear boundary, the nature of the dynamo must change in fully convective stars so that an $\alpha^2$ mechanism with a dynamo distributed throughout the convection zone would then be expected to generate and maintain magnetic activity \citep{kuker99alpha}.

Whether a change in the dynamo mechanism occurs as the \hbox{$\alpha$-effect} becomes dominant remains unclear.  With no shear boundary, the nature of the dynamo must change in fully convective stars so that an $\alpha^2$ mechanism with a dynamo distributed throughout the convection zone would then be expected to generate and maintain magnetic activity \citep{kuker99alpha}. Nevertheless, the mass at which stars become fully convective is unclear, since a change in the nature of dynamo is reasonably expected to lead to a break in activity-rotation indicators. Only recently has such a discontinuity been discovered \citep{mohanty03activity}, albeit at a much later spectral type (M9) than stellar structure models normally predict. These authors suggest that an increasingly neutral atmosphere could result in a drop of magnetic activity at this spectral type. Additionally, \citet{mullan01mdwarfs} have shown that the internal strength of the magnetic field can modify stellar structure significantly such that stars become fully convective at later spectral type than is predicted by standard evolution models.

While there is no evidence for a discontinuity in activity { indicators at early M spectral types,} other observations may be indicative of a change in the nature of dynamo activity. Although coronal X-rays tend to be saturated in the rapidly rotating regime, decreasing { optical} lightcurve amplitude for late K through to early-mid-M \citep{messina03spots} could be suggestive of uniformly distributed spots. This may be explained by distributed dynamo activity. Also, the model of \citep{kuker99alpha} indicates that the magnetic field is {\em steady}, and has a bipolar geometry, but with the axis inclined at 90\degs~to the rotation axis. While magnetic Zeeman imaging may be required to test the prediction of an inclined field, uniformly distributed spots and a steady state are indicated by similar distributions of starspots on the dM1.5 star HK Aqr \citep{barnes01mdwarfs, barnes04hkaqr} at three epochs over a period of 11 years.

\end{document}